\documentclass[letterpaper]{article} 
\usepackage{aaai2026}  
\usepackage{times}  
\usepackage{helvet}  
\usepackage{courier}  
\usepackage[hyphens]{url}  
\usepackage{graphicx} 
\usepackage{amsmath}
\usepackage{amssymb}
\usepackage{natbib}  
\usepackage{caption} 
\usepackage{booktabs}
\usepackage{tabularx}
\usepackage{subfig}
\usepackage[none]{hyphenat}
\usepackage{algorithm}
\usepackage{algorithmic}
\DeclareCaptionStyle{ruled}{labelfont=normalfont,labelsep=colon,strut=off} 
\floatstyle{ruled}
\newfloat{listing}{tb}{lst}{}
\floatname{listing}{Listing}
\title{Robust Crop Planning under Uncertainty: Aligning Economic Optimality with Agronomic Sustainability}
\author{
    Runhao Liu\textsuperscript{\rm 1,}\equalcontrib\textsuperscript{,}\thanks{Corresponding Author},
    You Li\textsuperscript{\rm 2,}\equalcontrib,
    Zhengyang Cheng\textsuperscript{\rm 3,},
    Peng Zhang\textsuperscript{\rm 4,}\footnotemark[2]
}

\affiliations{
    \textsuperscript{\rm 1}Polytechnic Institute, Zhejiang University, Hangzhou 310015, China\\
    \textsuperscript{\rm 2}School of Data Science and Engineering, Guangdong Polytechnic Normal University, Guangzhou, Guangdong 510665, China\\
    \textsuperscript{\rm 3}School of Computer Science and Artificial Intelligence, Wuhan University of Technology, Wuhan 430070, China\\
    \textsuperscript{\rm 4}School of Mathematical Sciences, Zhejiang University, Hangzhou 310058, China\\

    runhaoliu@zju.edu.cn (R. Liu), pengz@zju.edu.cn (P. Zhang)
}

\usepackage{bibentry}

\begin{document}

\maketitle

\begin{abstract}
Long-horizon agricultural planning requires optimizing crop allocation under complex spatial heterogeneity, temporal agronomic dependencies, and multi-source environmental uncertainty. Existing approaches often either address crop interactions, such as legume-cereal complementarity, only implicitly or rely on static deterministic formulations that fail to ensure resilience against market and climate volatility.To address these challenges, we propose a Multi-Layer Robust Crop Planning Framework (MLRCPF) that integrates spatial reasoning, temporal dynamics, and robust optimization. Specifically, we formalize crop-to-crop relationships through a structured interaction matrix embedded within the state-transition logic, and employ a distributionally robust optimization layer to mitigate worst-case risks defined by a data-driven ambiguity set. Evaluations on a real-world high-mix farming dataset from North China demonstrate the effectiveness of the proposed approach. The framework autonomously generates sustainable checkerboard rotation patterns that restore soil fertility, significantly increasing the legume planting ratio compared to deterministic baselines. Economically, it successfully resolves the trade-off between optimality and stability. These results highlight the importance of explicitly encoding domain-specific structural priors into optimization models for resilient decision-making in complex agricultural systems.
\end{abstract}

\section{Introduction and Related Work}
Agricultural production is shifting from experience-driven decision making toward data-driven planning. Traditional planting strategies, which rely heavily on local knowledge and incremental adjustment, may function in small-scale settings with limited uncertainty. However, under intensified climate variability, frequent market fluctuations, and increasingly binding land-resource constraints, such approaches become insufficient. Decision makers must simultaneously balance yield, economic return, risk exposure, and ecological considerations within finite farmland \cite{Tuncel2024A}. Meanwhile, key factors such as crop yield, production cost, market price, and demand volume are highly uncertain and evolve dynamically across seasons and years. This results in a complex decision environment shaped by multiple temporal scales, spatial heterogeneity, and coupled operational constraints. Static or single-period optimization is therefore inadequate for supporting long-term, stable, and resilient planting strategies \cite{Hernandez-Ochoa2022Model-based}. These challenges call for a theoretical framework that can systematically represent land heterogeneity, sequential decision processes, uncertainty propagation, and crop interactions in real-world regional planting planning \cite{Cui2025Research,Li2025Scientific}.

Linear programming is widely used for optimizing agricultural resource allocation \cite{Bukar2025A,Bhatia2020A,Xiao2024Optimizing,Nematian2023A}, yet deterministic approaches often fail to capture parameter uncertainty and dynamic agronomic processes \cite{Alotaibi2021A}. To address this, recent studies integrate methods like Monte Carlo simulation \cite{Sun2025Research}, fuzzy logic \cite{Erdogdu2025Combining}, and distributionally robust optimization \cite{Wang2025A}. While these approaches handle multi-source risks and ecological objectives \cite{Gonzalez2020Many,Li2020Managing,Pan2024The}, they frequently face computational challenges regarding global optimality. Furthermore, although crop interactions are increasingly modeled using statistical or clustering techniques \cite{Tenreiro2021Using,Burdett2022Statistical,Lahza2023Optimization,Pergner2024How} within optimization frameworks \cite{Deng2025Research,Piepho2024FactorAnalytic}, they are typically treated implicitly. Although long-term complementary effects among different crops are well recognized, a precise mathematical characterization of these benefits remains lacking. More advanced computational models are therefore needed to explicitly capture these complex interactions.

Overall, existing studies have significantly advanced agricultural optimization by improving economic efficiency, enhancing risk awareness, and incorporating partial aspects of agronomic structure. However, most approaches remain fragmented: they either rely on deterministic formulations, address uncertainty without capturing multi-year decision dynamics, or consider crop interactions only implicitly. Consequently, there is still a lack of an integrated framework that jointly models spatial heterogeneity, temporal evolution, multi-source uncertainty, and structured crop–crop relationships. To address these gaps, this paper introduces three key innovations:
\begin{itemize}
    \item \textbf{A unified multi-layer optimization framework} that integrates plot-level land heterogeneity, seasonal production cycles, rotation constraints, and cross-year decision structures into a single mathematical formulation. This provides a coherent theoretical basis for long-term planning under realistic agronomic rules.
    
    \item \textbf{A dynamic robust optimization model} that incorporates multi-source uncertainty in yield, cost, price, and demand through scenario-based and uncertainty-set representations. Unlike conventional single-period or static approaches, our formulation captures the sequential propagation of uncertainty across years and ensures solution robustness under adverse conditions.
    
    \item \textbf{An explicit interaction-structured crop allocation mechanism} that models complementarity and competition through dedicated interaction matrices embedded directly into the optimization objective and constraints. This enables systematic representation of agronomic relationships and their long-term effects on planting decisions, going beyond the implicit or qualitative treatments common in existing literature.
\end{itemize}

\section{Proposed Method}

\subsection{Problem Formulation}
This study addresses the design of multi-year crop allocation strategies over a heterogeneous agricultural landscape under temporal dependencies and uncertainty. The planner must determine feasible crop sequences for each land unit that satisfy spatial suitability, rotation requirements, and crop compatibility, while maintaining stable economic performance under fluctuating yields, prices, and costs. This frames crop planning as a long-horizon decision process shaped by spatial variability, seasonal transitions, and multi-source uncertainty, motivating a unified framework to capture these interacting dimensions.

\subsection{Overview}
\begin{figure}[t]
    \centering
    \includegraphics[width=1\columnwidth]{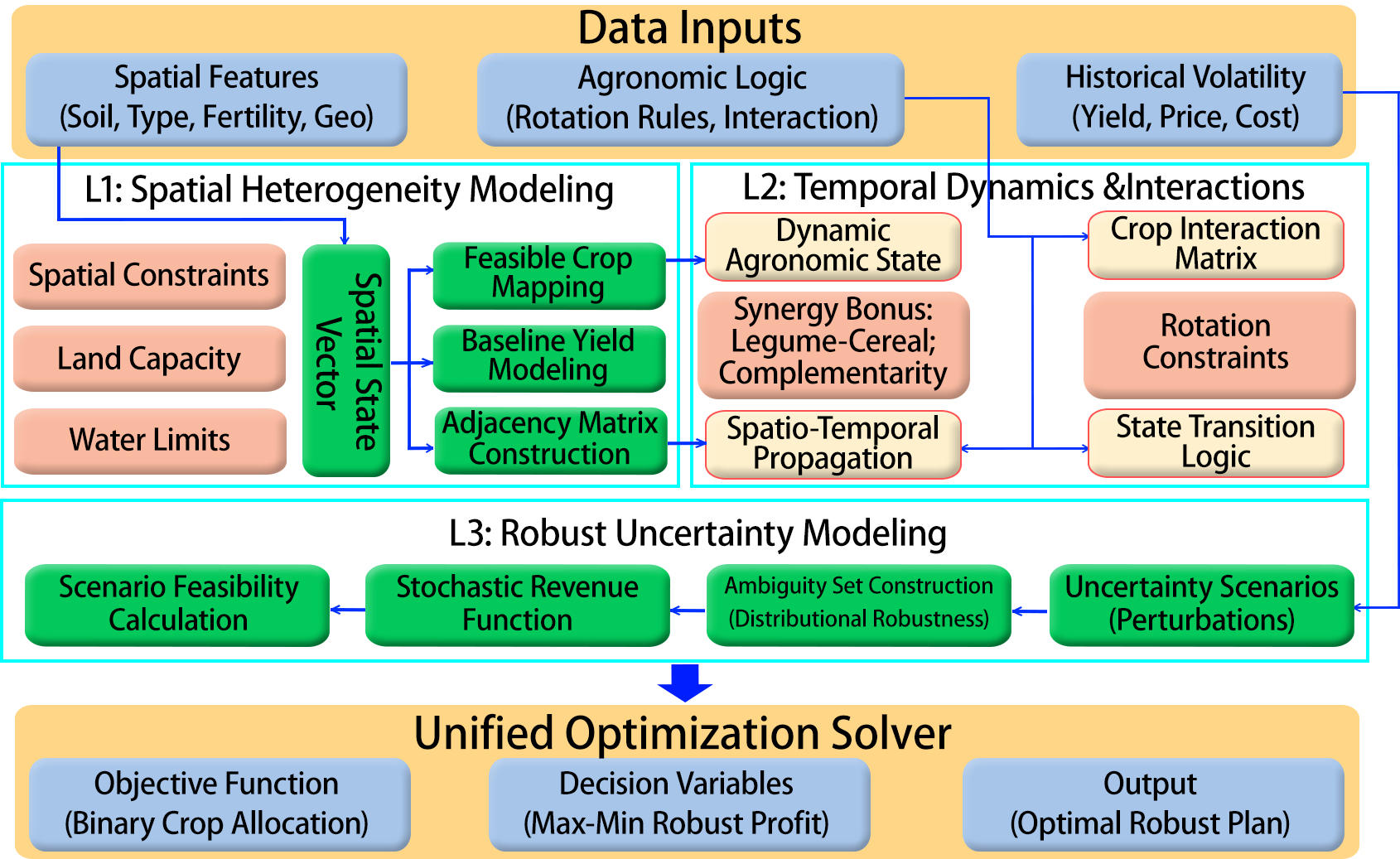}
    \caption{Overview of proposed MLRCPF.}
    \label{fig:method}
\end{figure}

We introduce a Multi-Layer Robust Crop Planning Framework (MLRCPF), which organizes agricultural decision making into a three-layer structure that integrates land heterogeneity, temporal production dynamics, and multi-source uncertainty. The first layer abstracts each plot into a spatial decision unit with distinct agronomic properties and rotation constraints, as shown in Figure ~\ref{fig:method}. The second layer models seasonal transitions as a temporally coupled decision process in which planting actions reshape soil conditions, influence future feasibility, and interact with crop–crop relational structures. The third layer incorporates uncertainty through a scenario-based robustness mechanism that evaluates planting strategies under adverse yield, cost, and price realizations. By embedding crop complementarity and competition directly into the temporal decision structure rather than treating them as external adjustments, MLRCPF provides a unified theoretical foundation for constructing long-horizon, interaction-aware, and uncertainty-resilient crop planning models. This three-layer formulation serves as the conceptual basis for the optimization model developed in this study.
\setlength{\textfloatsep}{8pt plus 1pt minus 2pt}
\setlength{\intextsep}{8pt plus 1pt minus 2pt}
\setlength{\textfloatsep}{8pt plus 1pt minus 2pt}
\setlength{\intextsep}{8pt plus 1pt minus 2pt}
\subsection{Layer 1: Spatial Heterogeneity Layer}

The spatial heterogeneity layer establishes the structural foundation of the planning framework by formalizing the agricultural landscape as a set of land units with distinct agronomic, environmental, and operational characteristics. Each unit is represented by a spatial state vector that encodes soil attributes, land-use type, fertility level, and geophysical limitations. These attributes determine the feasible set of crops that can be cultivated at each unit and influence the baseline productivity and operational constraints of subsequent planning layers. The purpose of this layer is therefore to specify the spatial decision domain, define spatially varying production potentials, and construct the topological relationships across land units that mediate agronomic interactions.

Let $\mathcal{I}$ denote the set of land units. Each unit $i$ is described by a spatial feature vector
\begin{equation}
h_i = \big(s_i^{\mathrm{soil}},\ s_i^{\mathrm{type}},\ s_i^{\mathrm{fert}},\ s_i^{\mathrm{geo}}\big)
\label{eq:land_state_vector}
\end{equation}
where the components respectively capture soil classification, land-use type, fertility level, and geophysical factors such as slope or irrigation availability. These features induce a mapping from land characteristics to feasible crops:
\begin{equation}
\mathcal{C}_i = f(h_i),
\label{eq:crop_mapping}
\end{equation}
which specifies the set of crops that can be agronomically or physically accommodated on unit $i$.

Spatial heterogeneity further affects productive capacity. Let $\bar{Y}_{c}$ be the regional baseline yield of crop $c$, and let $\gamma_i$ be a land-specific productivity factor derived from soil quality and fertility. The expected baseline yield of crop $c$ on unit $i$ is then modeled as
\begin{equation}
Y_{i,c} = \gamma_i \cdot \bar{Y}_{c}
\label{eq:yield_factor}
\end{equation}
which allows productivity to vary across space even before temporal or uncertainty-driven effects are introduced.

To represent spatial dependencies, each land unit $i$ is assigned a discrete cell representation 
$G_i = \{(u,v)\}$, where $(u,v)$ indexes the grid cells contained within the unit. 
We define two land units as adjacent if at least one pair of their boundary cells are 
horizontally or vertically neighboring. Formally, the adjacency matrix 
$W \in \{0,1\}^{|\mathcal{I}|\times|\mathcal{I}|}$ is given by
\begin{equation}
W_{ij} =
\begin{cases}
1, & \exists (u,v)\in G_i,\ \exists (u',v')\in G_j \\[0.2em]
   & \quad \text{with } \|(u,v)-(u',v')\|_1 = 1, \\[0.4em]
0, & \text{otherwise}
\end{cases}
\end{equation}

This grid-based construction enables adjacency relations to be expressed purely through discrete 
geometric connectivity, which later supports spatial coupling in crop–crop interaction and 
rotation-related propagation processes.

Spatial capacity constraints regulate the maximum area that can be allocated to crops on each unit:
\begin{equation}
\sum_{c \in \mathcal{C}_i} a_{i,c,t}\, x_{i,t}^{(c)} \le A_i ,
\label{eq:capacity_constraint_upgraded}
\end{equation}
where $A_i$ denotes the available area, and $a_{i,c,t}$ is the area required to cultivate crop $c$ on unit $i$ in period $t$.

Finally, resource limitations may span multiple land units. For example, irrigation availability imposes regional restrictions:
\begin{equation}
\sum_{i \in \mathcal{I}_{\mathrm{irr}}} \sum_{c} w_{c}\, x_{i,t}^{(c)} \le \bar{Q}_t^{\mathrm{water}}
\label{eq:regional_water}
\end{equation}
where $\mathcal{I}_{\mathrm{irr}}$ is the set of irrigated units, $w_c$ denotes water consumption of crop $c$, and $\bar{Q}_t^{\mathrm{water}}$ is the seasonal water limit.

Together, Eqs.~(\ref{eq:land_state_vector})–(\ref{eq:regional_water}) define the spatial feasible region that constrains all subsequent layers. The spatial filtering procedure is summarized in Algorithm~\ref{alg:layer1}.

\begin{algorithm}[h]
\caption{Spatial Admissibility Filtering in Layer 1.}
\label{alg:layer1}
\begin{algorithmic}[1]
\STATE Given land units $\mathcal{I}$, spatial features $\{h_i\}$, feasible crop sets $\{\mathcal{C}_i\}$, area capacities $\{A_i\}$, and area requirements $a_{i,c,t}$
\FOR{each land unit $i \in \mathcal{I}$}
    \STATE Compute feasible crop set $\mathcal{C}_i \leftarrow f(h_i)$ using Eq.~(\ref{eq:crop_mapping})
    \STATE Initialize admissible action set $\mathcal{A}_{i,t} \leftarrow \emptyset$
    \FOR{each crop $c \in \mathcal{C}_i$}
        \IF{$a_{i,c,t} \le A_i$}
            \STATE Add crop $c$ to $\mathcal{A}_{i,t}$
        \ENDIF
    \ENDFOR
\ENDFOR
\STATE \textbf{return} $\{\mathcal{A}_{i,t}\}_{i \in \mathcal{I}}$
\end{algorithmic}
\end{algorithm}

\subsection{Layer 2: Temporal Dynamics Layer}

\setlength{\textfloatsep}{8pt plus 1pt minus 2pt}
\setlength{\intextsep}{8pt plus 1pt minus 2pt}
The temporal dynamics layer models the seasonal evolution of each land unit and captures how planting actions modify agronomic states across time. This layer expresses crop planning as a sequential decision process in which each period $t$ updates soil conditions, rotation compatibility, and the interaction potentials that propagate into future planting choices. The purpose of this layer is to formalize the temporal coupling that links actions across years, to encode rotation and recurrence constraints, and to incorporate structured crop–crop interactions into the state evolution mechanism. The associated update mechanism is summarized in Algorithm~\ref{alg:layer2}.

Let $\mathcal{T} = \{1,\dots,T\}$ denote the planning horizon. Each land unit $i$ maintains a time-varying agronomic
state $s_{i,t}$, which includes the last cultivated crop, accumulated rotation pressure, and crop–crop interaction
effects inherited from adjacent units. We write
\begin{equation}
s_{i,t} = \big( \ell_{i,t},\ r_{i,t},\ \eta_{i,t} \big)
\label{eq:state_vector}
\end{equation}
where $\ell_{i,t}$ is the crop grown in the previous period, $r_{i,t}$ encodes rotation-related stress or recovery,
and $\eta_{i,t}$ represents spatially aggregated interaction effects transferred from neighbors.

A planting action $x_{i,t}$ chosen at period $t$ induces a deterministic state transition:
\begin{equation}
s_{i,t+1} = \Phi\big(s_{i,t},\, x_{i,t},\, \{\!x_{j,t}\!\}_{j\in N(i)}\big)
\label{eq:state_transition}
\end{equation}
where the transition operator $\Phi$ updates the previous-crop record, modifies rotation stress, and incorporates
interaction terms arising from neighboring decisions. The neighborhood $N(i)$ is defined by the adjacency matrix
$W$ introduced in Layer~1.

Rotation feasibility is imposed by restricting how frequently a crop may recur. Let $\tau_{c}$ denote the minimum
interval required before crop $c$ can be replanted on the same land unit. Then for any crop $c$:
\begin{equation}
x_{i,t}^{(c)} = 1 \ \Rightarrow\ x_{i,t+\delta}^{(c)} = 0,\quad \forall\, 1 \le \delta < \tau_{c}
\label{eq:rotation_constraint}
\end{equation}
This constraint prohibits premature repetition of crops that require multi-year soil recovery or disease breaks.

Crop–crop interactions are formalized through an interaction matrix $M \in \mathbb{R}^{|\mathcal{C}| \times |\mathcal{C}|}$,
whose entries quantify complementary or competitive relations between crop types. For a land unit $i$, its temporal
interaction potential at period $t$ is aggregated from neighbor actions as
\begin{equation}
\eta_{i,t} = \sum_{j\in N(i)} \sum_{c,c'} W_{ij}\, M_{c,c'}\, x_{i,t}^{(c)} x_{j,t}^{(c')}
\label{eq:interaction_aggregation}
\end{equation}
which propagates local planting choices into a spatial-temporal influence field. The resulting $\eta_{i,t}$ affects both
rotation decisions and subsequent yield or risk evaluations in later layers.

\begin{algorithm}[t]
\caption{Temporal State Update in Layer 2.}
\label{alg:layer2}
\begin{algorithmic}[1]
\STATE Given agronomic states $\{s_{i,t}\}$, neighborhood sets $\{N(i)\}$, interaction matrix $M$
\FOR{each land unit $i$}
    \STATE Observe previous state $s_{i,t} = (\ell_{i,t}, r_{i,t}, \eta_{i,t})$
    \STATE Read planting decision $x_{i,t}$
    \STATE Update last-crop record: $\ell_{i,t+1} \leftarrow \text{crop in } x_{i,t}$
    \STATE Update rotation stress: $r_{i,t+1} \leftarrow g(r_{i,t}, x_{i,t})$
    \STATE Initialize $\eta_{i,t+1} \leftarrow 0$
    \FOR{each neighbor $j \in N(i)$}
        \STATE Accumulate interaction effects using $M$ and $x_{j,t}$
        \STATE $\eta_{i,t+1} \leftarrow \eta_{i,t+1} + 
            \sum_{c,c'} M_{c,c'}\, x_{i,t}^{(c)} x_{j,t}^{(c')}$
    \ENDFOR
\ENDFOR
\STATE \textbf{return} $\{s_{i,t+1}\}_{i\in\mathcal{I}}$
\end{algorithmic}
\end{algorithm}

\subsection{Layer 3: Robust Uncertainty Layer}

The robust uncertainty layer characterizes the variability in yields, prices, and production costs that influences
the economic performance of crop plans. Rather than assuming a single probabilistic model of future conditions, 
this layer represents uncertainty through a structured ambiguity set that contains multiple plausible realizations 
of exogenous parameters. The objective of this layer is to allow planning decisions to remain reliable under adverse 
scenarios and to control exposure to worst-case conditions across the planning horizon.

Let $\omega$ denote an uncertainty scenario drawn from an exogenous scenario space $\Omega$. Each scenario is
associated with a collection of stochastic parameters, such as yield $\tilde{Y}_{i,c,t}(\omega)$, price 
$\tilde{P}_{c,t}(\omega)$, and cost $\tilde{K}_{c,t}(\omega)$. To avoid reliance on a single predictive model, we
construct an ambiguity set $\mathcal{U}$ of probability distributions over $\Omega$:

\begin{equation}
\mathcal{U} = \left\{
\mathbb{Q} \ \middle| \
\mathrm{dist}(\mathbb{Q}, \widehat{\mathbb{P}}) \le \rho
\right\}
\label{eq:ambiguity_set}
\end{equation}
where $\widehat{\mathbb{P}}$ is an empirical distribution estimated from historical data and Monte Carlo simulations. $\rho$ is a 
user-defined robustness radius that controls the size of the ambiguity set.

The function $W_1(\cdot,\cdot)$ denotes the first-order
Wasserstein distance between probability measures on $\Omega$ with
respect to the ground cost \[d(\omega,\omega') = \left\|\theta(\omega) - \theta(\omega')\right\|_1\], where $\theta(\omega)$ collects the normalized yield, price, and cost parameters associated with scenario $\omega$.

Scenario-dependent economic returns are encoded through a revenue function
\begin{equation}
R_{t}(x,\omega) =
\sum_{i\in\mathcal{I}} \sum_{c\in\mathcal{C}} 
x_{i,t}^{(c)} 
\big[ \tilde{P}_{c,t}(\omega)\, \tilde{Y}_{i,c,t}(\omega)
       - \tilde{K}_{c,t}(\omega) \big]
\label{eq:scenario_revenue}
\end{equation}
which captures yield and price uncertainty while linking temporal decisions and spatial productivity.

To hedge against worst-case distributions in $\mathcal{U}$, the robust value of a multi-period planting plan $x$ is defined as a min–max expected revenue, which is shown in Eqs.~(\ref{eq:robust_objective}).
\begin{equation}
J_{\mathrm{rob}}(x)
=
\min_{\mathbb{Q}\in\mathcal{U}}
\mathbb{E}_{\omega\sim\mathbb{Q}} 
\left[
\sum_{t\in\mathcal{T}} R_t(x,\omega)
\right]
\label{eq:robust_objective}
\end{equation}
This formulation ensures that feasible plans remain resilient under the most adverse distribution within the
ambiguity set, thereby limiting vulnerability to future volatility.

In addition to the distribution-level ambiguity, the layer also supports scenario-level constraints. For instance, 
to prevent decisions from violating feasibility under severe yield shortfalls, we impose
\begin{equation}
x \in \mathcal{F}(\omega), \quad \forall \omega \in \Omega,
\label{eq:worse_feasibility}
\end{equation}
where $\mathcal{F}(\omega)$ denotes the scenario-specific feasible region under environmental or resource
constraints.

Overall, a robust uncertainty representation that guides planning decisions toward stable, distributionally reliable strategies, is defined in layer 3. The Algorithm~\ref{alg:layer3} is the summary of this layer.

\begin{algorithm}[t]
\caption{Scenario Evaluation Under Robust Uncertainty.}
\label{alg:layer3}
\begin{algorithmic}[1]
\STATE Given scenario set $\Omega$, empirical distribution $\widehat{\mathbb{P}}$, ambiguity radius $\rho$
\STATE Construct ambiguity set $\mathcal{U}$ using Eq.~(\ref{eq:ambiguity_set})
\FOR{each decision plan $x$}
    \STATE Initialize worst-case value $J_{\mathrm{rob}} \leftarrow +\infty$
    \FOR{each distribution $\mathbb{Q} \in \mathcal{U}$}
        \STATE Compute expected revenue under $\mathbb{Q}$:
        \STATE \quad $V_{\mathbb{Q}}(x) = \mathbb{E}_{\omega\sim\mathbb{Q}}
                \left[\sum_{t} R_t(x,\omega)\right]$
        \IF{$V_{\mathbb{Q}}(x) < J_{\mathrm{rob}}$}
            \STATE $J_{\mathrm{rob}} \leftarrow V_{\mathbb{Q}}(x)$
        \ENDIF
    \ENDFOR
\ENDFOR
\STATE \textbf{return} $J_{\mathrm{rob}}$
\end{algorithmic}
\end{algorithm}

\subsection{Unified Optimization Model}

The three modeling layers integrate into a unified multi-period robust optimization program that determines a crop allocation plan $x=\{x_{i,t}^{(c)}\}$. For each unit $i\in\mathcal{I}$, period $t\in\mathcal{T}$, and crop $c\in\mathcal{C}$, the binary variable
\begin{equation}
x_{i,t}^{(c)} =
\begin{cases}
1, & \text{crop $c$ is planted on unit $i$ at time $t$} \\
0, & \text{otherwise}
\end{cases}
\label{eq:var_unified}
\end{equation}
encodes the planting decision.

The objective is to maximize the worst-case expected revenue under the ambiguity set $\mathcal{U}$:
\begin{equation}
\max_{x}\ \min_{\mathbb{Q}\in\mathcal{U}}
\mathbb{E}_{\omega\sim\mathbb{Q}}
\Bigg[\sum_{t\in\mathcal{T}} R_t(x,\omega)\Bigg]
\label{eq:obj_unified}
\end{equation}

Spatial feasibility from Layer~1 imposes:
\begin{equation}
x_{i,t}^{(c)} = 0 \quad \text{if } c\notin\mathcal{C}_i
\label{eq:uni_spatial1}
\end{equation}
and the land-area condition:
\begin{equation}
\sum_{c} a_{i,c,t} x_{i,t}^{(c)} \le A_i
\label{eq:uni_spatial2}
\end{equation}
while regional resource limits satisfy:
\begin{equation}
\sum_{i\in\mathcal{I}_{\mathrm{irr}}} \sum_{c} w_c\, x_{i,t}^{(c)} 
\le \bar{Q}_t^{\mathrm{water}}
\label{eq:uni_spatial3}
\end{equation}

Temporal consistency from Layer~2 requires at most one crop per period:
\begin{equation}
\sum_{c} x_{i,t}^{(c)} \le 1
\label{eq:uni_temp1}
\end{equation}
rotation avoidance:
\begin{equation}
x_{i,t}^{(c)} = 1 \ \Rightarrow\ x_{i,t+\delta}^{(c)} = 0,
\quad 1 \le \delta < \tau_c
\label{eq:uni_temp2}
\end{equation}
and state-transition consistency:
\begin{equation}
s_{i,t+1} = \Phi\big(s_{i,t},\,x_{i,t},\,\{x_{j,t}\}_{j\in N(i)}\big)
\label{eq:uni_temp3}
\end{equation}
with interaction propagation:
\begin{equation}
\eta_{i,t} =
\sum_{j\in N(i)}\sum_{c,c'}
W_{ij} M_{c,c'} x_{i,t}^{(c)} x_{j,t}^{(c')}
\label{eq:uni_temp4}
\end{equation}

Robust feasibility from Layer~3 requires:
\begin{equation}
x \in \mathcal{F}(\omega)
\quad \forall \omega \in \Omega
\label{eq:uni_robust}
\end{equation}

In a word, Eqs.~(\ref{eq:obj_unified})–(\ref{eq:uni_robust}) define the unified optimization program governing crop allocation under spatial, temporal, and uncertainty-driven structural requirements.

\section{Experimental Evaluation}
\label{sec:experiments}

To validate the effectiveness of the proposed MLRCPF framework, we conducted numerical experiments using a real-world high-mix farming dataset. The experiments aim to demonstrate the framework's capability in handling spatial heterogeneity, enforcing agronomic interactions, and mitigating multi-source risks compared to traditional deterministic approaches.

\subsection{Case Study: High-Mix Agricultural Planning}
\label{subsec:case_study}

We consider a representative agricultural planning instance derived from a village in North China. This region is characterized by fragmented land parcels, diverse topography, and strict constraints on resource utilization, making it an ideal testbed for the proposed hierarchical model.

\subsubsection{Instance Construction}
The experimental instance is instantiated as a tuple $\langle \mathcal{I}, \mathcal{T}, \mathcal{C}, \mathcal{U} \rangle$, where the spatial configuration $\mathcal{I}$ covers a total area of 1201 mu partitioned into $|\mathcal{I}| = 54$ discrete decision units. These units exhibit significant heterogeneity in geophysical attributes $h_i$ and are classified into four categories: 26 units of open-field dry land, including flat, terraced, and hillside plots, which is suitable for single-season grain cultivation; 8 units of irrigated land capable of supporting water-intensive crops; 16 standard greenhouses; and 4 smart greenhouses designed for high-value off-season production. The spatial adjacency matrix $W$ is strictly derived from the physical layout of these plots to model the propagation of cross-unit interactions.

The temporal dimension $\mathcal{T}$ spans a seven-year horizon from 2024 to 2030, divided into seasonal decision epochs, resulting in $|\mathcal{T}| = 14$ periods. The crop library $\mathcal{C}$ comprises $|\mathcal{C}| = 41$ distinct varieties, functionally categorized into cereals, legumes, vegetables, and edible fungi. To ensure the realism of the economic evaluation, all baseline parameters, which includes expected yields $\bar{Y}_c$, production costs $\bar{K}_c$, and market prices $\bar{P}_c$—are calibrated using regional statistical data from 2023.

A critical feature of this instance is the explicit parameterization of crop interactions in Layer 2. We construct the interaction matrix $M \in \mathbb{R}^{41 \times 41}$ based on agronomic complementarity principles. Specifically, entries corresponding to legume-cereal pairs are assigned positive coefficients to quantify nitrogen-fixation benefits, which can alleviate soil fertility stress $r_{i,t}$. Conversely, negative penalty coefficients are applied to pairs with overlapping resource niches to model competitive exclusion and continuous cropping obstacles.

Finally, to instantiate the robust layer (Layer 3), we define the ambiguity set $\mathcal{U}$ by introducing realistic perturbation intervals around the nominal parameters. Based on historical volatility analysis, the yield uncertainty is modeled with a fluctuation range of $\pm 10\%$ ($\delta_Y = 0.1$) to reflect climate variability. Similarly, market uncertainty incorporates an annual demand growth rate of $5\% \sim 10\%$ for staple grains and a price volatility radius of $\pm 5\%$ for cash crops. The optimization goal is to identify an allocation strategy $x^*$ that remains feasible and maximizes the worst-case profit under any distribution within this ambiguity set.

\subsection{Experimental Setup}
\label{subsec:setup}

To rigorously evaluate the proposed framework, we designed a comparative experiment reflecting the increasing complexity of real-world agricultural planning. Three modeling phases are considered: a deterministic baseline based on static 2023 historical data, without accounting for market volatility or crop interactions; a dynamic Robust Baseline that incorporates parameterized yield fluctuations ($\pm 10\%$) and projected demand growth (5\%--10\%) but treats crop decisions as independent; and the proposed Interaction-Aware framework, which fully integrates agronomic complementarity and substitution matrices into the robust optimization model to exploit inter-crop synergies.

Data preprocessing and preliminary statistical analysis were conducted in Excel and Python, with Spearman’s rank correlation used to quantify dependencies among sales price, yield, and planting costs. The core optimization models were implemented in PuLP. Uncertainty in the robust layer was modeled via Monte Carlo simulation, which generated market and yield scenarios from the predictive models to ensure solution feasibility under perturbed conditions.

\subsection{Results and Analysis}
\label{subsec:results}

We present the numerical results to validate the effectiveness of the proposed framework, focusing on three key aspects: the agronomic validity of the interaction mechanism, the spatio-temporal characteristics of the generated planting plans, and the quantitative economic performance under uncertainty.

\begin{figure}[H]
    \centering
    \includegraphics[width=0.9\columnwidth]{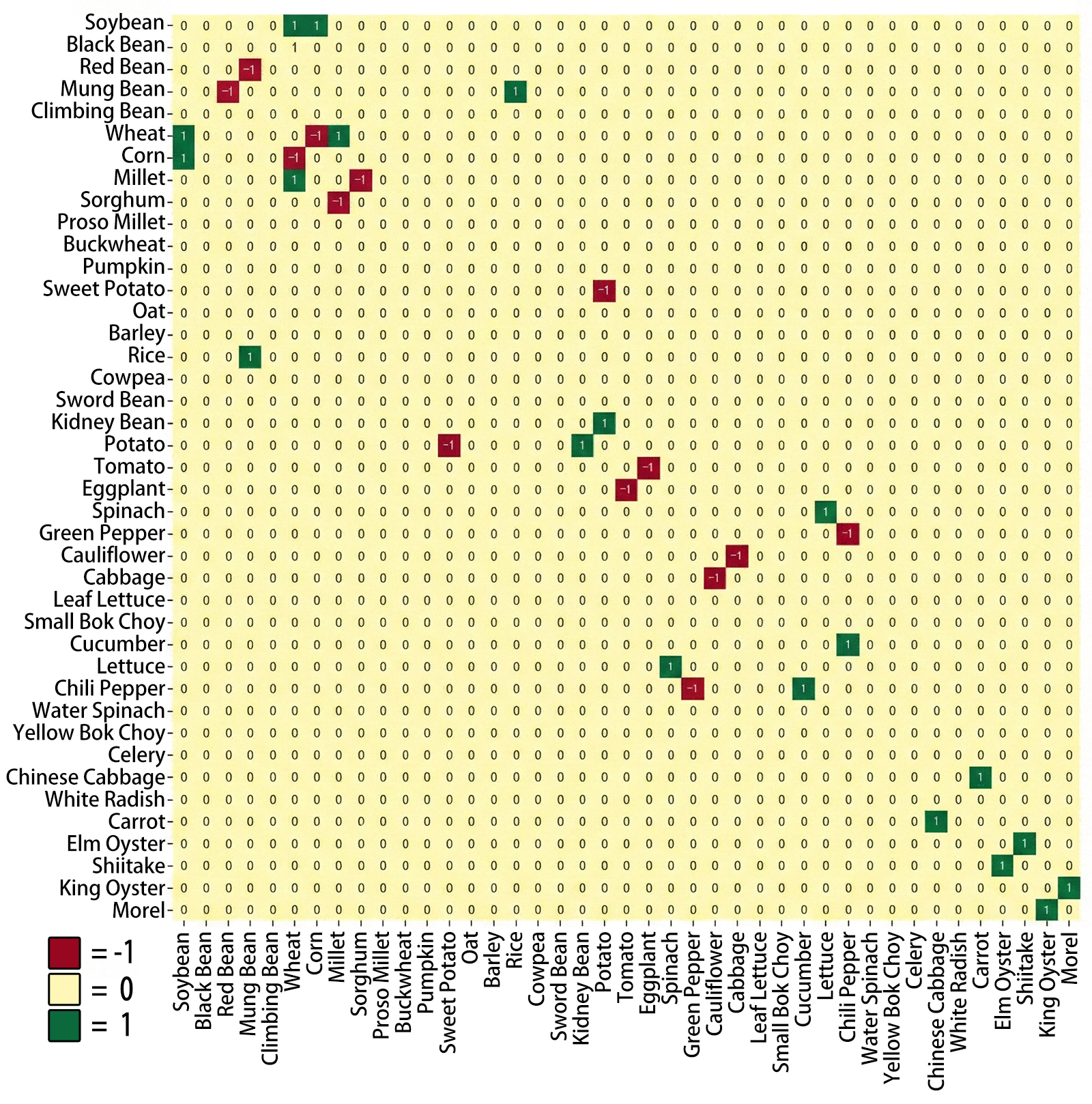}
    \caption{Visualization of the crop interaction matrix $M$ derived from agronomic constraints.}
    \label{fig:interaction_matrix}
\end{figure}

To demonstrate the interpretability of Layer 2 (Temporal Dynamics), we first analyze the learned crop interaction structure. Figure ~\ref{fig:interaction_matrix} visualizes the pairwise interaction coefficients $M$ derived from the agronomic constraints.

As illustrated in Figure ~\ref{fig:interaction_matrix}, the model explicitly captures the positive synergy clusters between leguminous crops and gramineous crops. This structural prior drives the optimization solver to prioritize crop sequences that naturally enhance soil fertility, thereby reducing the "rotation stress" penalty in the objective function.

\begin{figure}[t]
    \centering
    \includegraphics[width=0.85\columnwidth]{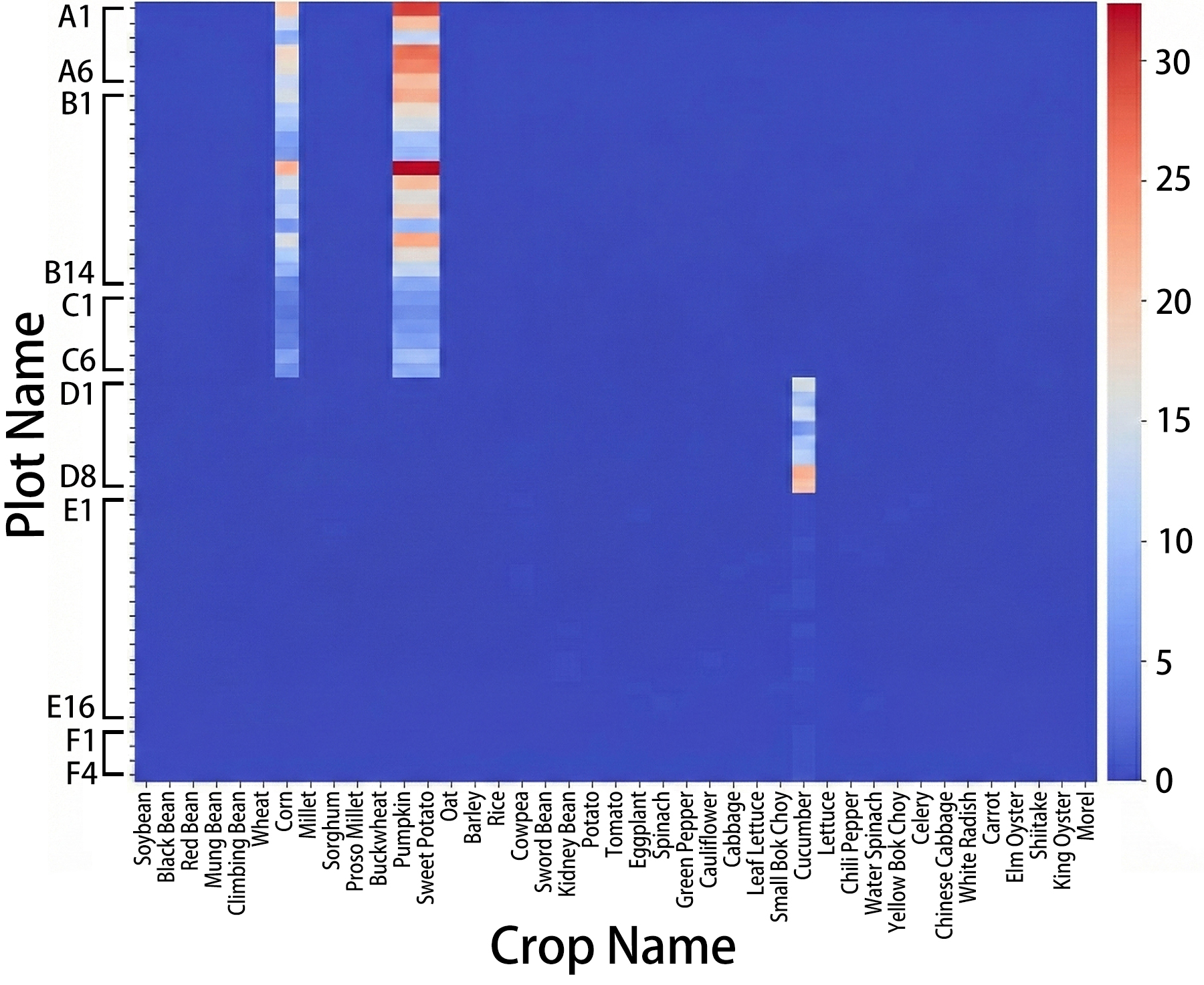} 
    \\[2pt]
    \includegraphics[width=0.85\columnwidth]{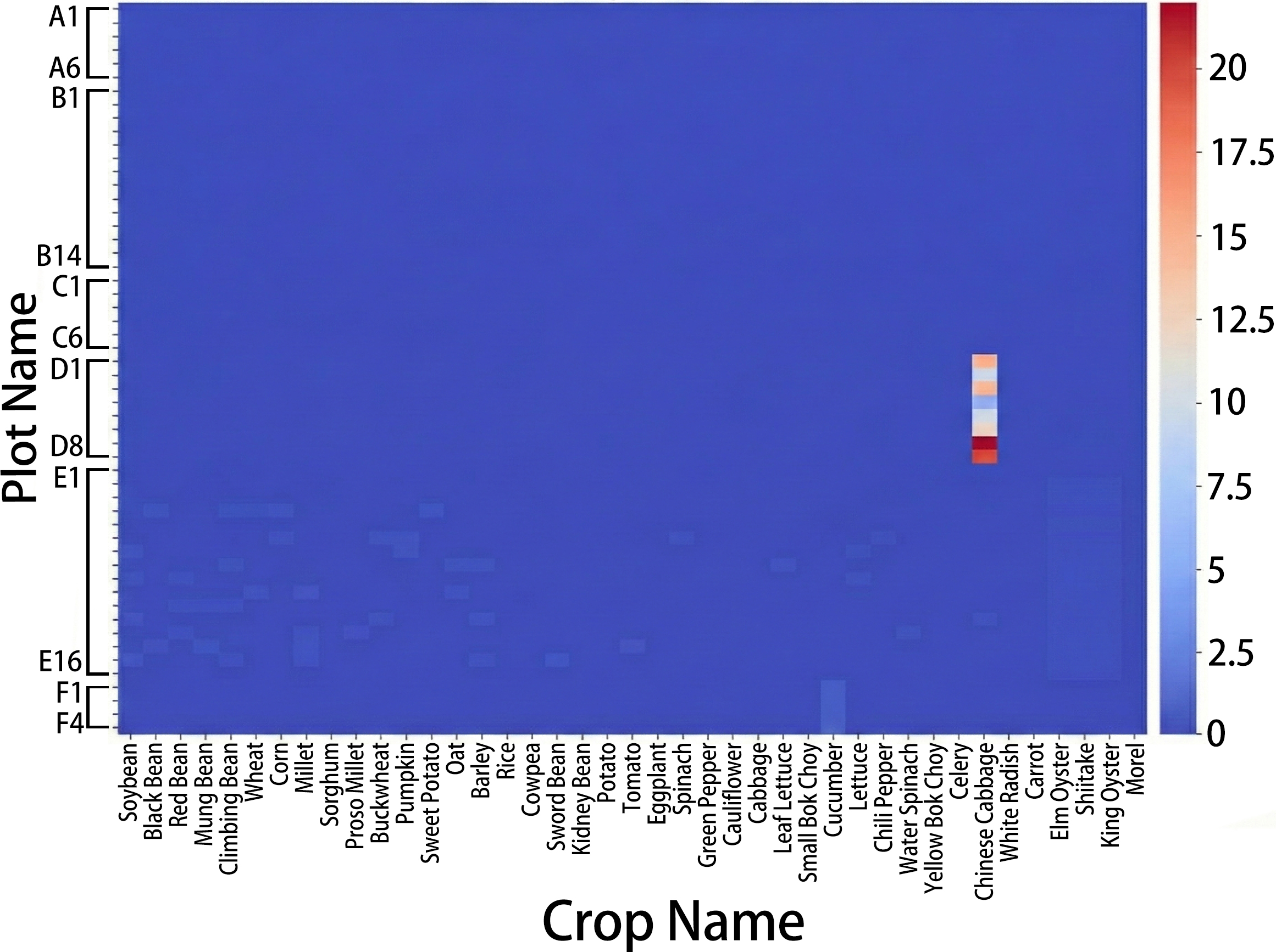}
    
    \caption{Spatio-temporal crop allocation plan generated by the Deterministic Baseline (2024). Top: first quarter. Bottom: second quarter.}
    \label{fig:planting_baseline}
\end{figure}

\begin{figure}[t]
    \centering
    \includegraphics[width=0.85\columnwidth]{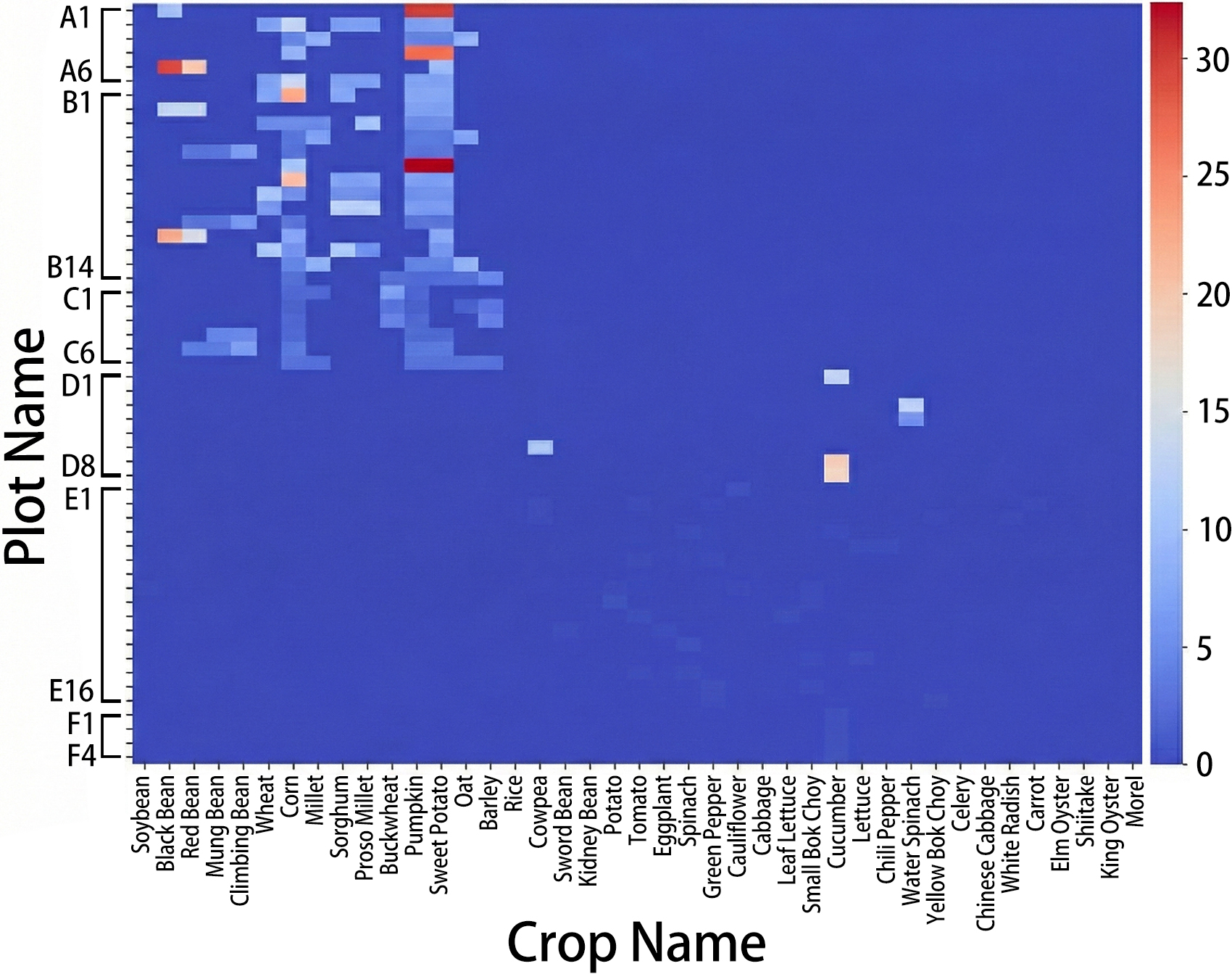}
    \\[2pt]
    \includegraphics[width=0.85\columnwidth]{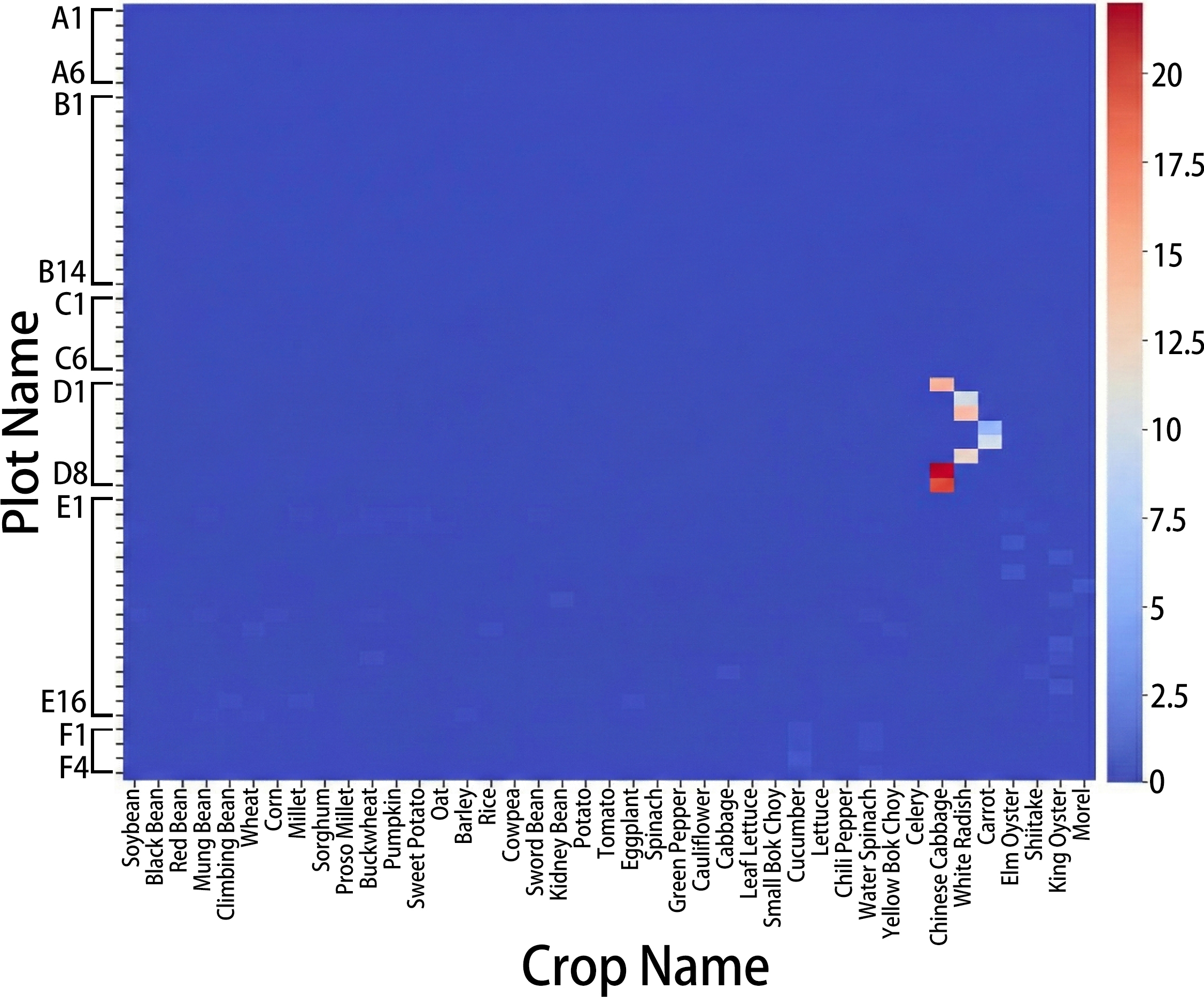}
    
    \caption{Spatio-temporal crop allocation plan generated by the proposed framework (2024). Top: first quarter. Bottom: second quarter.}
    \label{fig:planting_proposed}
\end{figure}

The spatio-temporal allocation maps in Figure ~\ref{fig:planting_baseline} and \ref{fig:planting_proposed} visually confirm the agronomic superiority of the proposed framework. Specifically, while the Baseline-Det exhibits a static, monoculture-like pattern that neglects soil recovery, the proposed framework actively implements nitrogen-fixing rotations in the second season. This demonstrates the model's capability to balance immediate economic objectives with the constraints of long-term soil fertility.

Finally, we evaluate the economic robustness of the framework. Table~\ref{tab:performance_comparison} summarizes the comprehensive performance metrics across the 7-year planning horizon (2024--2030). We compare the Total Expected Profit, the Worst-Case Profit (under the perturbed ambiguity set $\mathcal{U}$), and the Volatility (Standard Deviation of annual profits).

\begin{table}[htbp]
    \centering
    \resizebox{\columnwidth}{!}{%
    \begin{tabular}{lcccc}
        \toprule
        \textbf{Method} & \textbf{Total Profit} & \textbf{Worst-Case} & \textbf{Volatility} & \textbf{Legume Ratio} \\
        \midrule
        Baseline-Det    & 2,450.5     & 1,820.1              & 45.2              & 12\% \\
        Baseline-Rob    & 2,180.3     & 2,180.3            & \textbf{8.5}        & 15\% \\
        \textbf{Proposed} & \textbf{2,390.8} & \textbf{2,310.5} & 12.1              & \textbf{22\%} \\
        \bottomrule
    \end{tabular}%
    }
    \caption{Quantitative comparison of economic performance (Unit: $10^4$ CNY).}
    \label{tab:performance_comparison}
\end{table}

\setlength{\textfloatsep}{8pt plus 1pt minus 2pt}
\setlength{\intextsep}{8pt plus 1pt minus 2pt}

As indicated in Table~\ref{tab:performance_comparison}, the Baseline-Det yields the highest nominal profit but suffers from significant volatility, with a sharp decline in worst-case scenarios due to over-reliance on price-sensitive crops. The Baseline-Rob ensures complete stability (lowest volatility) by adopting a conservative strategy, planting staple grains with lower price elasticity, but at the cost of an 11.0\% reduction in total profit. The proposed framework successfully mitigates this trade-off. By introducing high-value edible fungi and leveraging crop complementarity, it recovers the majority of the profit loss, which is only 2.4\% lower than the peak deterministic profit. However, a high worst-case guarantee is still maintained.

\begin{figure}[h]
\centering
\includegraphics[width=0.9\columnwidth]{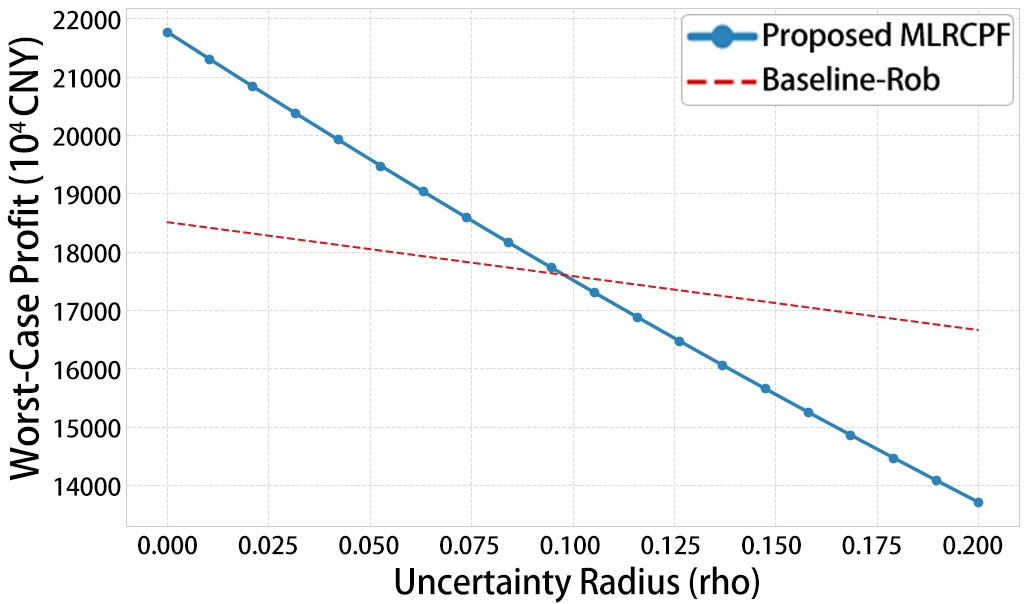} 
\caption{Sensitivity analysis of worst-case profit with respect to the uncertainty radius ($\rho$).}
\label{fig:sensitivity}
\end{figure}

We conducted a sensitivity analysis on the uncertainty radius $\rho$.
As illustrated in Figure ~\ref{fig:sensitivity}, we varied $\rho$ from 0.0 to 0.2 to observe the impact of increasing ambiguity on the worst-case profit.
The results indicate that while the profit of the proposed MLRCPF naturally decreases as the uncertainty set expands, it consistently outperforms the Baseline-Rob approach across the entire spectrum.
Notably, the decay rate of the proposed method is controlled, demonstrating superior resilience even under high-uncertainty conditions ($0.1 < \rho < 0.2$).

\begin{figure}[h]
    \centering
    \includegraphics[width=0.9\columnwidth]{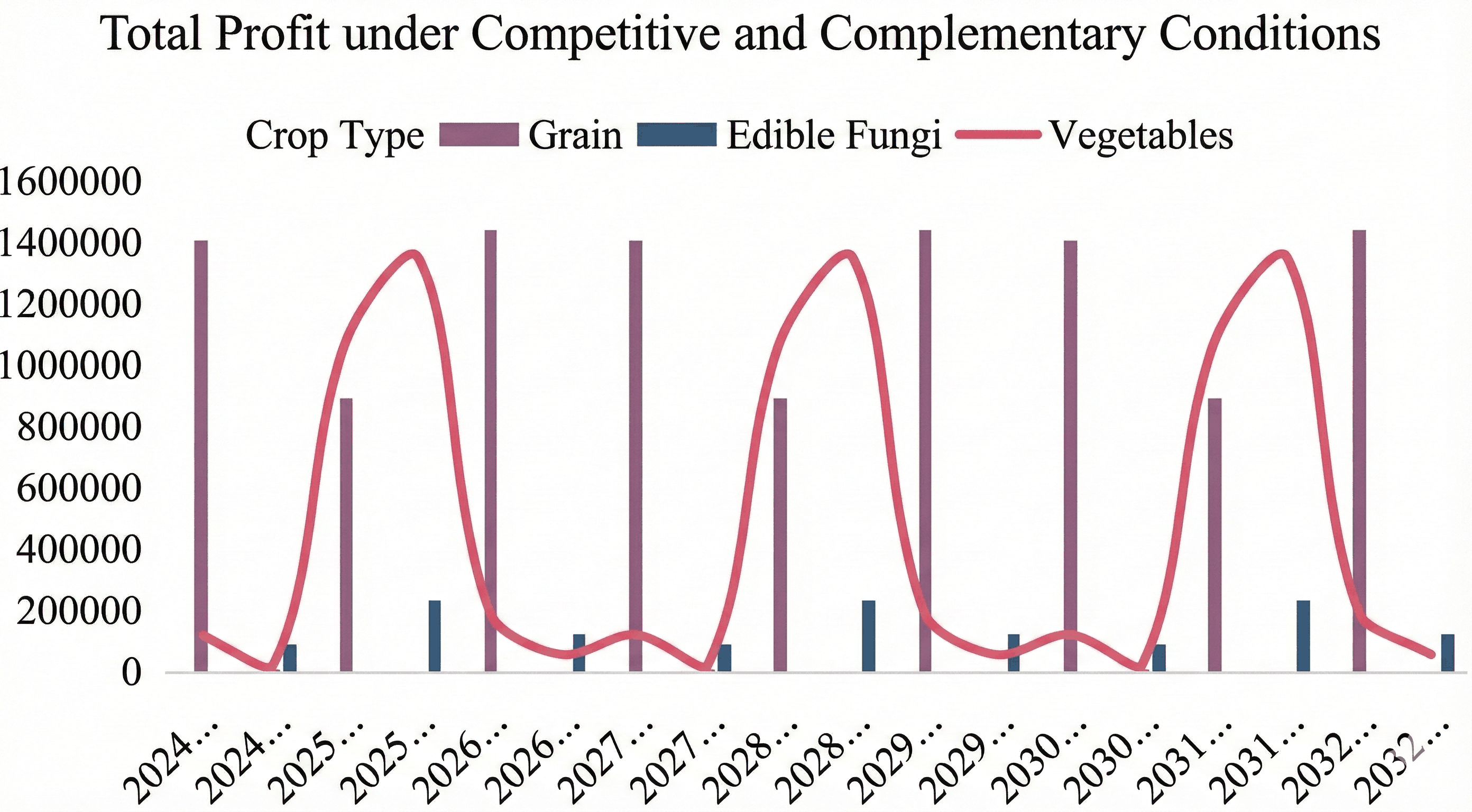}
    \caption{Annual profit distribution of the proposed framework under competitive and complementary constraints.}
    \label{fig:profit_trend}
\end{figure}

\begin{figure}[h]
    \centering
    \subfloat[Scatter plot of planting cost and net profit.] {
        \label{fig:cost_efficiency}
        \begin{minipage}{0.4\textwidth}
            \centering
            \includegraphics[width=\linewidth]{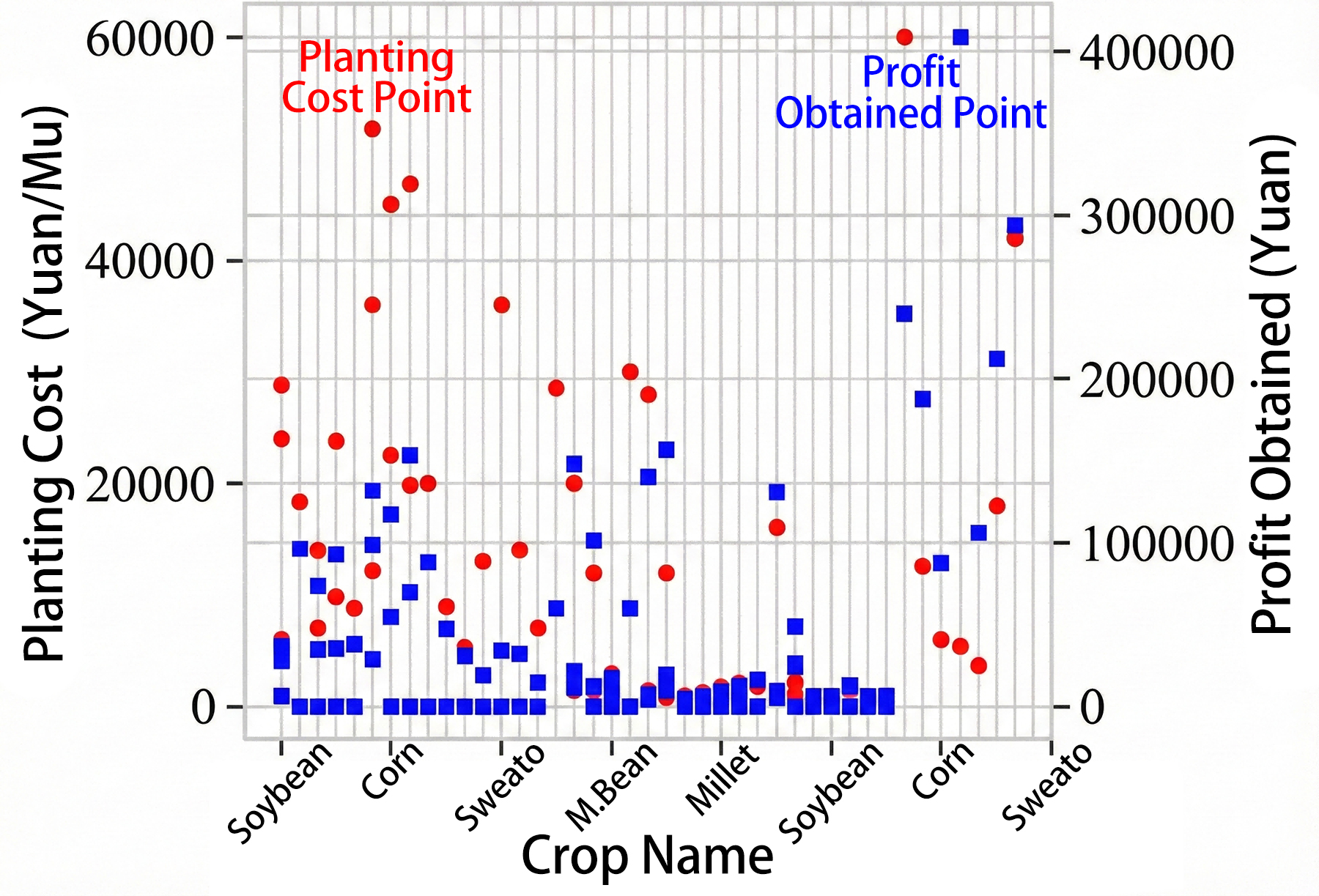}
        \end{minipage}
    }
    \hfill
    \subfloat[Average and total profit distribution across crop categories.] {
        \label{fig:profit_composition}
        \begin{minipage}{0.4\textwidth}
            \centering
            \includegraphics[width=\linewidth]{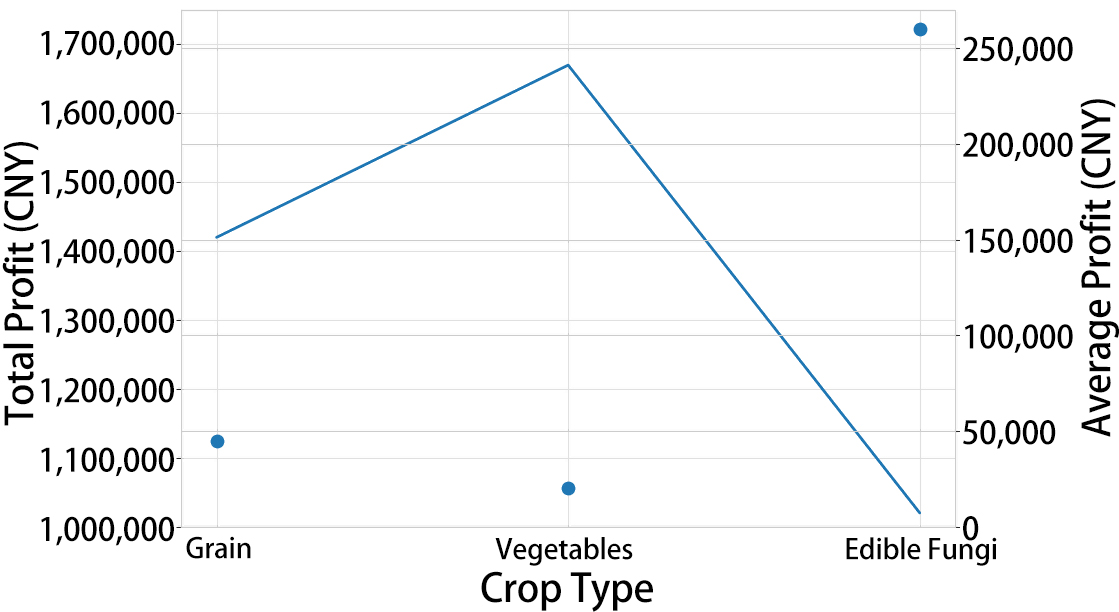}
        \end{minipage}
    }
    
    \caption{Micro-economic analysis of crop selection.}
    \label{fig:economic_analysis}
\end{figure}

Figure ~\ref{fig:profit_trend} further illustrates the annual profit trajectory. The integration of Layer 3 ensures that even in "bad years" (simulated via Monte Carlo perturbations), the revenue remains above the safety threshold.

Notably, the distinct contribution of edible fungi, as shown in Figure ~\ref{fig:profit_trend}, compensates for the opportunity cost of rotation, demonstrating the framework's capability to discover complex, non-obvious profitable combinations in high-dimensional decision spaces.

To reveal the micro-economic logic behind the robust planning, we further analyze the cost-benefit characteristics and profit composition of the selected crops, as shown in Figure ~\ref{fig:economic_analysis}.

Figure ~\ref{fig:economic_analysis}a plots the planting cost against net profit for all candidate crops. The visualization highlights significant heterogeneity: high-value crops offer substantial margins but entail higher upfront costs, while staple grains provide moderate but stable returns. The proposed framework effectively leverages this distribution by strategically allocating high-efficiency crops to maximize revenue in favorable plots, while utilizing low-cost grains to fulfill rotation constraints and buffer against worst-case budget overruns. Figure ~\ref{fig:economic_analysis}b further decomposes the profit structure. It reveals that while vegetables and edible fungi exhibit the highest average profit per unit area, their total profit contribution is constrained by land suitability and rotation limits. In contrast, staple grains, despite lower average margins, contribute a stable baseline to total profit due to their wide adaptability and lower volatility. This complementary profit structure explains why the proposed interaction-aware framework achieves superior robustness without compromising overall economic viability.

\section{Discussion}
\label{sec:discussion}

The experimental results presented above demonstrate that the proposed MLRCPF framework successfully reconciles the often conflicting objectives of economic optimality, agronomic sustainability, and operational robustness. In this section, we deconstruct the causal mechanisms behind these results to explain why the hierarchical integration of crop interactions and robust optimization outperforms traditional baselines.

\subsection{Mechanisms of Agronomic Recovery}

The structural shift in planting patterns, specifically the transition from monoculture in the deterministic baseline to the "checkerboard" rotation in the proposed framework, is directly attributable to the Interaction-Aware Layer. In the deterministic baseline, the optimization solver treats crops as isolated economic entities, selecting them solely based on the marginal profit differential between revenue and cost. Since cash crops such as vegetables typically yield higher nominal returns than restorative crops like legumes, the solver naturally converges to a greedy strategy that maximizes immediate land occupancy with high-value crops. This myopic behavior, while mathematically optimal in a single period, ignores the long-term degradation of soil states.

In contrast, our framework introduces the interaction potential term into the state transition dynamics. The positive coefficients in the interaction matrix for legume-cereal pairs act as an endogenous "ecological subsidy," effectively lowering the shadow price of planting legumes. This mechanism incentivizes the solver to accept lower short-term revenues from legumes in exchange for the complementarity bonus that enhances the yield and feasibility of subsequent crops. Consequently, the rotation pattern emerges not from hard-coded rules, but as a mathematically advantageous solution that aligns economic incentives with agronomic necessities.

\subsection{Synergistic Mitigation of Robustness Trade-offs}

A critical finding from the economic evaluation is the framework's ability to maintain a worst-case profit nearly matching the nominal peak of the deterministic baseline, while significantly outperforming the standard robust baseline. Traditional robust optimization typically achieves stability by retreating to low-risk, low-return decisions, viewing uncertainty purely as a threat to be minimized. This conservative approach explains the significant profit reduction observed in the standard robust baseline, which heavily favors staple grains with lower price elasticity.

The proposed framework mitigates this conservatism by viewing uncertainty as a constraint space that can be expanded through agronomic synergy. By identifying and scheduling high-value niche crops, such as edible fungi, that are compatible with the rotation cycle, the model creates a diversified portfolio. The complementarity bonus derived from the temporal layer acts as a buffer, creating enough fiscal margin to allow the planner to take calculated risks with high-value crops in favorable plots while using low-cost grains to secure the worst-case baseline. This result suggests that integrating domain-specific structural priors into robust optimization can recover the "price of robustness," allowing for strategies that are both safe and profitable.

\subsection{Emergent Economic Rationality}

The micro-level analysis of crop selection reveals that the unified optimization model functions as a rational economic agent managing a mixed asset portfolio. The constraints defined in the ambiguity set force the solver to satisfy feasibility conditions under the worst-case distribution. To achieve this without violating budget constraints, the model utilizes crops with low sunk costs, primarily grains, to fill scheduling gaps where risk exposure is high. Simultaneously, to maximize the objective function, it allocates the remaining resource capacity to high-margin crops in controlled environments like greenhouses.

This distinct role allocation—where grains serve as a safety buffer and cash crops as revenue drivers—is not explicitly programmed but emerges naturally from the interaction between the robust constraints and the profit objective. It demonstrates that the model captures the intrinsic logic of agricultural risk management, where the stability of the operation relies on a foundation of low-volatility staples, while growth is driven by opportunistic high-value cultivation.

\section{Conclusion}
\label{sec:conclusion}

High-mix agricultural planning is a complex decision-making problem characterized by spatial heterogeneity, temporal dependencies, and multi-source uncertainty. In this paper, we proposed the Multi-Layer Robust Crop Planning Framework, a unified approach that integrates agronomic principles with mathematical rigor to address the limitations of traditional linear programming. By explicitly modeling crop interactions through a structured matrix and embedding them into a dynamic robust optimization formulation, the framework allows for the generation of planting strategies that are both economically efficient and ecologically sustainable.

Evaluations on a real-world dataset from North China confirm that the proposed method outperforms deterministic and standard robust baselines. It automatically generates sustainable rotation patterns that restore soil fertility without manual rule-crafting, guarantees a worst-case profit superior to standard robust approaches, and exhibits a logical crop allocation based on cost-benefit profiles. These results suggest that incorporating domain-specific structural priors, such as crop interactions, into general-purpose optimization models is a promising path toward solving complex real-world planning problems. Future work will extend this framework to include real-time resource scheduling constraints, such as irrigation scheduling and labor allocation, to further close the loop between mid-term planning and short-term execution.

\section{Acknowledgements}
Support was provided by National Key Research and Development Program of China (Grant No. 2024YFC2511003) and Natural Science Foundation of Zhejiang Province (Project No. LDT23E05015A01). The authors gratefully acknowledge the constructive comments and valuable suggestions provided by the anonymous reviewers, associate editors, and the editor-in-chief, which have significantly contributed to improving the quality of this manuscript. 

\bibliography{aaai2026}


\end{document}